\documentclass{article}

\def\abstract#1{\vskip 7mm 
        \begin{center}{\large Abstract}\par \smallskip
                \begin{minipage}[c]{12cm}
                        \small #1
                \end{minipage}
        \end{center}
}
\def\title#1{\begin{center}{\Large\bf #1}\end{center}}
\def\author#1{\vskip 5mm \begin{center}{#1}\end{center}}
\def\address#1{\begin{center}{\it #1}\end{center}}
\makeatletter
\@ifundefined{lesssim}{}{}
\@ifundefined{gtrsim}{}{}
\def\vereq#1#2{\lower3pt\vbox{\baselineskip1.5pt \lineskip1.5pt
\ialign{$\m@th#1\hfill##\hfil$\crcr#2\crcr\sim\crcr}}}
\makeatother

\begin{document}

\title{%
Born-Again Braneworld
}
\author{%
Sugumi Kanno,\footnote{E-mail:kanno@phys.h.kyoto-u.ac.jp}
Misao Sasaki,\footnote{E-mail:misao@vega.ess.sci.osaka-u.ac.jp}
and Jiro Soda\footnote{E-mail:jiro@phys.h.kyoto-u.ac.jp}
}
\address{%
$^1$Graduate School of Human and Environmental Studies, Kyoto University,
 Kyoto 606-8501, Japan\\
$^2$Department of Earth and Space Science, Graduate School of Science,
Osaka University, \\ Toyonaka 560-0043, Japan\\
$^3$Department of Fundamental Sciences, FIHS, Kyoto University, 
Kyoto 606-8501, Japan
}

\abstract{
We propose a cosmological braneworld scenario in which two branes
collide and emerge as reborn branes with signs of tensions
opposite to the original tensions of respective branes.
In this scenario, the branes are assumed to be
inflating. However,  the whole dynamics is different 
from the usual inflation due to the non-trivial dynamics
of the radion field. Transforming the conformal frame to
the Einstein frame, this born-again scenario resembles 
the pre-big-bang scenario. Thus our scenario has features
of both inflation and pre-big-bang scenarios.
In particular, the gravitational
waves produced from vacuum fluctuations will have a very
blue spectrum, while the inflaton field will give rise to
a standard scale-invariant spectrum.      
}

\section{Introduction}

The inflationary universe scenario is a natural solution to
fundamental problems of the big-bang model such as horizon 
problem. However, it is not a unique choice.  A universe with an era
of contraction is also a possibility.
The pre-big-bang scenario is a realization of such a case in the
superstring context\cite{prebigbang}.
Unfortunately, however, the pre-big-bang scenario suffers from
a singularity problem which cannot be solved without 
understanding the stringy non-perturbative effects.

One of the remarkable features of superstring theory is the existence
of extra dimensions. Conventionally, the extra dimensions are considered
to be compactified to a small compact space of the Planck scale. 
However, recent revolutionary progress in string theory has lead to
the brane-world picture \cite{braneworld}.
In this talk, we consider a system of two branes having tensions
of opposite signs, with the intermediate
spacetime (bulk) described by an anti-de Sitter space (AdS$_5$) \cite{RS1}. 
One of the branes is assumed to be our universe, and
there exists an inflaton field which leads to inflation.
The other brane is assumed to be vacuum
but with a non-zero cosmological constant.

Assuming slow roll of the inflaton field,
we may regard both branes as vacuum (de Sitter) branes.
So far, mostly the static de-Sitter two-brane system is considered in
the cosmological context. However, it is now well-known that
a static de-Sitter two-brane system is unstable.  
We therefore investigate the non-trivial radion dynamics and
dwell on its cosmological consequences~\cite{Kanno:2002py}. 

\section{Radion Dynamics}

Most inflationary models are based on slow-roll inflation
which have a sufficiently flat potential.
In this section, we consider dynamics of branes with vacuum energy as  
a first order approximation of a slow-roll inflation model.
Qualitative features of the brane cosmology can be understood by
this simplified vacuum brane model.

We take the matter Lagrangeans to be ${\cal L}^\oplus=-\delta\sigma^\oplus$ 
and  
${\cal L}^\ominus=-\delta\sigma^\ominus$. 
The effective action on the positive ($\oplus$) tension brane for this setup 
reads~\cite{kanno1}
 (see also \cite{wiseman,kanno2,kanno3})
\begin{eqnarray}
S_\oplus = {\ell\over 2 \kappa^2} \int d^4 x \sqrt{-h}
	\left[
	\Psi R
	- {3\over2(1-\Psi)}
	\Psi^{|\alpha}
	\Psi_{|\alpha} \right]
	-\delta\sigma^\oplus \int d^4 x \sqrt{-h} 
	-\delta\sigma^\ominus \int d^4 x \sqrt{-h} 
	\left(1-\Psi\right)^2. 
	\label{A:action-vcm}
\end{eqnarray}
Since our theory is a scalar-tensor type theory, we call
this original action the Jordan-frame effective action.
 In order to discuss the dynamics of radion, it is convenient to 
move to the Einstein frame in which the action takes the canonical
Einstein-scalar form. 
Applying a conformal transformation 
$h_{\mu\nu}=\frac{1}{\Psi}g_{\mu\nu}$ and 
introducing a new field $
\eta=-\log\left|\frac{\sqrt{1-\Psi}-1}{\sqrt{1-\Psi}+1}\right| \ ,
\label{A:field}
$
we obtain the Einstein-frame effective action 
\begin{eqnarray}
S_\oplus &=&{\ell\over 2 \kappa^2} \int d^4 x \sqrt{-g}
	\left[R(g)-\frac{3}{2}
	\nabla^\alpha\eta
	\nabla_\alpha\eta\right]
	-\int d^4x\sqrt{-g}~V(\eta)  \ ,
	\label{EH-action}
\end{eqnarray}
where $\nabla$ denotes the covariant derivative with respect to
 the metric $g_{\mu\nu}$ and the radion potential now takes the form,
\begin{equation}
V(\eta) = \delta\sigma^\oplus\left[~\cosh^4\frac{\eta}{2}
	+\beta\sinh^4\frac{\eta}{2}~\right], \quad 
	\beta=\frac{\delta\sigma^\ominus}{\delta\sigma^\oplus} \ .
	\label{potential}
\end{equation}

We can also start from the effective action 
on the negative ($\ominus$) tension brane to obtain the same Einstein-frame 
effective action.
By a conformal transformation $f_{\mu\nu}=\frac{1}{\Phi}g_{\mu\nu}$ 
and introducing a new field $
\eta=-\log\left|\frac{\sqrt{\Phi+1}-1}{\sqrt{\Phi+1}+1}\right|  \ ,
\label{B:field}
$
we also arrive at Eq.~(\ref{EH-action}). 

We note that the two branes are infinitely separated when $\eta=0~(\Psi=1)$
and they collide when $\eta=\infty~(\Psi=0)$. 
For definiteness, let us assume $\delta\sigma^\oplus>0$. 
If $\delta\sigma^\oplus + \delta\sigma^\ominus>0$, $\Psi$ will
move towards unity, i.e., the branes will move away from each other.
While, if $\delta\sigma^\oplus+\delta\sigma^\ominus<0$,
the potential has a maximum at $\Psi_c=1+1/\beta$, and the behavior
depends on whether $\Psi>\Psi_c$ or $\Psi<\Psi_c$. 
If $\Psi>\Psi_c$, the branes will become infinitely separated.
If $\Psi<\Psi_c$, the branes will approach to each other
and eventually collide.  From the 5-dimensional point of view,
this is surely a singularity where the spacetime degenerates
to 4-dimensions. However, as far as observers on the branes
are concerned, nothing seems to go wrong.
In fact, the action~(\ref{A:action-vcm}) is
well-defined even in the limit $\Psi\to0$.
Let us assume that $\Psi$ smoothly become negative after collision.
Then replacing $\Psi$ as $\Psi\to -\tilde{\Psi}$ in the
action (\ref{A:action-vcm}), we find
\begin{eqnarray}
-S_\oplus &=& {\ell\over 2\kappa^2} \int d^4 x \sqrt{-h}
	\left[~
	\tilde{\Psi}R(h)
	+\frac{3}{2}\frac{1}{1+\tilde{\Psi}}
	\tilde{\Psi}^{|\alpha}\tilde{\Psi}_{|\alpha}~
	\right]
	+\int d^4 x \sqrt{-h}
	(-{\cal L}^\oplus) \nonumber \\
&&	+\int d^4 x \sqrt{-h} 
	\left(1+\tilde{\Psi}\right)^2
	({-\cal L}^\ominus) \ .
\end{eqnarray}
This is the same as the effective action on the negative tension brane
\begin{eqnarray}
S_\ominus&=&{\ell\over 2 \kappa^2} \int d^4 x \sqrt{-f} 
	\left[ \Phi R(f) + {3 \over 2(1+\Phi )} 
     	\Phi^{;\alpha} \Phi_{;\alpha} \right]  \nonumber\\
&&\!\!\!\!	+\int d^4 x \sqrt{-f} {\cal L}^\ominus
     	+\int d^4 x \sqrt{-f} {\cal L}^\oplus(1+\Phi)^2
     	\ .  
      	\label{B:action} 
\end{eqnarray}
 except for the overall change of sign and
the associated change of signs in front of the matter Lagrangeans.
One can interpret this fact as follows. After collision,
the positive tension brane turns into a negative tension brane
together with the sign change of the matter Lagrangean, and
vice versa for the initially negative tension brane.
This implies that, if we live on either of the branes,
our world transmutes into quite a different world and 
so does ourselves without much damage to the world.
 That is, we are born again!

\section{Born-Again Braneworld}

If our world had been initially a positive tension brane,
we would be now on the negative tension brane.
However, this case contradicts with observation.
Therefore we take the position that we were initially on the
negative tension brane before the collision. 

Let us first investigate the cosmological evolution of the
negative tension brane in the original Jordan frame.
We consider the spatially isotropic and homogeneous metric on
the brane,
\begin{eqnarray}
ds^2=-dt^2+a^2(t)\gamma_{ij}dx^idx^j\,,
\label{mtrc:cosmology}
\end{eqnarray}
where $a(t)$ is the scale factor and $\gamma_{ij}$ is the metric of a
maximally symmetric 3-space with comoving curvature $K=0$, $\pm1$.
The field equations on the negative tension brane give
\begin{eqnarray}
\dot{H}-\frac{K}{a^2}=-2\left(H^2+\frac{K}{a^2}\right)
-\frac{2\kappa^2}{3\ell}\delta\sigma^\ominus \ .
\label{B:ij2} 
\end{eqnarray}
Integrating this equation, we obtain the Friedmann equation 
with a dark radiation,
\begin{eqnarray}
H^2+\frac{K}{a^2}
=-\frac{\kappa^2}{3\ell}\delta\sigma^\ominus+\frac{C}{a^4} \ .
\label{friedmann}
\end{eqnarray}
We also find the relation between the radion and the dark radiation,
\begin{eqnarray}
&&\frac{\kappa^2\delta\sigma^\ominus}{3\ell}\frac{1+\Phi}{\Phi}
	\left[1+\frac{(1+\Phi)}{\beta}\right]  
	-H\frac{\dot{\Phi}}{\Phi}
	-\frac{1}{4}\frac{1}{1+\Phi}\frac{\dot{\Phi}^2}{\Phi}
	=\frac{C}{a^4}   \ . 
\label{relation}
\end{eqnarray}
This gives, in particular, the relation between the
initial conditions of the radion and the sign of the dark 
radiation.

To realize the born-again braneworld scenario,
we consider a case of colliding branes. For simplicity,
we assume $K=0$.  We can see numerically that $\Phi$ passes through
zero smoothly and approaches $-1$, i.e.,
the reborn branes will be eventually infinitely separated.

Let us analyze this collision. 
We denote the Hubble constant at the time of collision $t=t_c$ by $H_c$. 
Applying Eq.~(\ref{relation}) to the vicinity of the time of collision,
we find
\begin{eqnarray}
\Phi=-2(1-\sqrt{\gamma})H_c(t-t_c)\,;
\quad\gamma=1-\frac{H_*^2}{H_c^2}\left(1+\frac{1}{\beta}\right)\ ,
\end{eqnarray}
where $H_*^2=(\kappa^2/3\ell)(-\delta\sigma^\ominus)$. 
As expected, $\Phi$ behaves perfectly smoothly around
the time of collision. The brane geometry is, of course,
perfectly regular as well. In fact, the Friedmann
equation (\ref{friedmann}) continues to hold without
a hint of collision.

Now we transform these quantities into the Einstein frame.
Since $\Phi\to-1$ eventually, we may regard our present
universe to be described by the Einstein frame.
The relation between the Einstein frame and the Jordan frame is
\begin{eqnarray}
ds^2_E&=&-dt_E^2 + b^2 (t_E) \delta_{ij} dx^i dx^j \nonumber\\
	&=&|\Phi| \left[-dt_J^2+a(t_J)^2\delta_{ij}dx^idx^j\right],
\end{eqnarray}
where we attached the subscripts $E$ and $J$ to the time coordinates
to denote the cosmic time in the Einstein frame and the Jordan frame,
respectively.
Thus we have
$b=\sqrt{|\Phi|}\,a, \quad dt_E = \sqrt{|\Phi|}\,dt_J\,.
\label{conftrans}$
Therefore, the Hubble parameter in the Einstein frame 
behaves in the vicinity of collision as
\begin{eqnarray}
\frac{\dot{b}(t_E)}{b(t_E)}
=\frac{1}{3t_E} 
+\frac{H_c}{\left(3(1-\sqrt{\gamma})H_c|t_E|\right)^{1/3}}\,,
\label{Ehubble}
\end{eqnarray}
where the collision time in the Einstein frame is set to be $t_E=0$.

We note that in the Einstein frame, the universe is contracting
rapidly just before the collision and the Hubble parameter 
diverges to minus infinitely at collision.
Then the universe is reborn with an infinitely large Hubble parameter,
which looks like a big-bang singularity. Thus, since there exists
no singularity in the Jordan frame, the pre-big-bang phase and the
post-big-bang phase in the Einstein frame is successfully
connected. That is, our scenario is indeed a successful realization
of the pre-big-bang scenario in the context of the braneworld.

\section{Observational Implication}

 As we can see from Eq.~(\ref{friedmann}), the universe will rapidly
 converge to the quasi-de-Sitter regime, while the radion can 
 vary as far as the relation (\ref{relation}) is satisfied. 
 In the Jordan frame, as the metric couples with the radion,
 the non-trivial evolution of the radion field affects the perturbations.
 This possibility discriminate our model from the usual inflationary 
 scenario. On the other hand, the inflaton does not couple directly
with the radion field. Hence, the inflaton fluctuations are expected
to give adiabatic fluctuations with a flat spectrum. 
 This feature is an advantage compared with the pre-big-bang model.

\subsection{Radion fluctuation}

 To study the behavior of the radion fluctuations,
 it is convenient to work in the Einstein frame. 
The action for the curvature perturbation ${\cal R}$
on the $\delta\eta=0$ (i.e., radion-comoving) slice reads
\begin{eqnarray}
S={1\over2}\int d\eta\, d^3x\,z^2
\left[{\cal R}_c'{}^2-{\cal R}_c^{\,|i}{\cal R}_{c\,|i}\right]\,,
\label{ptb:action}
\end{eqnarray}
where ${\cal H}=b'/b$ and
\begin{eqnarray}
{\cal R}_c={\cal R}-{\cal H}{\delta\eta\over\eta'}\,,
\quad
z=\sqrt{3\ell\over2\kappa^2}\,{b\eta'\over{\cal H}}\,.
\label{Rcdef}
\end{eqnarray}

As the background behaves as $b \sim (-\tau)^{1/2} $,
${\cal H}\sim(2\tau)^{-1}$ and $\eta' \sim (-\tau)^{-1}$, 
we have $z\propto b$, and
the positive frequency modes for the adiabatic vacuum is given by
\begin{equation}
   {\cal R}_{c,k}\sim\sqrt{\pi\kappa^2\over6H_*\ell}\,
H_0^{(1)} (-k\tau ) \,,
\label{bessel}
\end{equation}
where we have normalized $b$ as $b=|H_*\tau|^{1/2}$.
Then we have
\begin{eqnarray}
\left\langle{\cal R}_c^2\right\rangle_k
\equiv{k^3\over 2\pi^2}P(k)
={k^3\over 2\pi^2}|{\cal R}_{c,k}|^2
\sim {k^3\over H_*M_{pl}^2}\,,
\label{curvamp}
\end{eqnarray}
where $M_{pl}^2=\kappa^2/\ell$. Thus the spectrum is very blue.
If we define the spectral index by $P(k)\propto k^{n-4}$,
this implies $n=4$.

\subsection{Gravitational waves}

Next, consider the tensor perturbations:
\begin{equation}
ds^2 = b^2 (\tau) \left[
       -d\tau^2 + (\delta_{ij} + h_{ij}) dx^i dx^j \right] 
\label{mtrc:tnsr-ptb}
\end{equation}
where $h_{ij}$ satisfy the transverse-traceless conditions, 
 $h_{ij}{}^{,j} = h^{i}{}_i =0$. 
As for the gravitational tensor perturbations, we have
\begin{equation}
    h_k^{\prime\prime} + 2{\cal H}h_k'+k^2h_k
    =0 \ ,
\label{tnsr:schrodinger-typ}
\end{equation}
where $h_k$ is the amplitude of $h_{ij}$.
Since ${\cal H}\sim (2\tau)^{-1}$,
$h_k$ has approximately the same spectrum as that for ${\cal R}_c$,
including the magnitude.
In particular, the spectral index for the gravitational waves is also
 $n=4$ (with the spectral index defined by $P_h(k)\propto k^{n-4}$ as in
the case of scalar curvature perturbation; for the tensor perturbation,
the conventional definition is $n_T=n-1$).
 Provided that inflation ends right after collision,
this gives a sufficiently
blue spectrum that can amplify $\Omega_g$ by several order of 
magnitudes or more on small scales as compared to conventional
inflation models. Thus, there arises a possibility that
it may be detected by a space laser interferometer for
low frequency gravitational waves such as LISA~\cite{detector}.
  
\subsection{Inflaton perturbation}

To investigate the inflaton perturbation rigorously,
one needs to introduce an inflaton field explicitly and
consider a system of equations fully coupled with
the radion and the metric perturbation. The estimation
shows that the effect of the metric perturbation
induced by radion fluctuations on the inflaton perturbation
 is small.  Hence, the inflaton fluctuations will have a standard
scale-invariant spectrum.

\section{Conclusion}

 In this paper, we proposed a scenario in which two branes collide
and are reborn as new branes, called the born-again braneworld scenario.
 Our model has the features of both inflationary and pre-big-bang scenarios. 
In the original frame, which we call the Jordan frame since
gravity on the brane is described by a scalar-tensor type
theory, the brane universe is assumed to be inflating due to an
  inflaton potential. 
 While, the radion, which represents the distance between the branes
and which acts as a gravitational scalar on the branes,
 has non-trivial dynamics and theses vacuum branes can collide and 
    pass through smoothly.
After collision, it is found that 
the positive tension and the negative tension 
 branes exchange their role.  Then, they move away from each other,
and the radion becomes trivial after a sufficient lapse of time.
The gravity on the originally negative tension
brane (whose tension becomes positive after collision) will then
approach the conventional Einstein theory except for 
tiny Kaluza-Klein corrections. 

One can also consider the cosmological evolution of the branes in
the Einstein frame. Note that the two frames are indistinguishable
at present if our universe is on the positive tension brane after
collision. In the Einstein frame, the brane universe is contracting
before the collision and one encounters a singularity 
at the collision point. This resembles the pre-big-bang scenario.
Thus our scenario may be regarded as a non-singular realization
of the pre-big-bang scenario in the braneworld context.   
 
As our braneworld is inflating and the inflaton has essentially
no coupling with the radion field, 
the adiabatic density perturbation with a flat spectrum is 
naturally realized.
  While, as the collision of branes mimics the pre-big-bang scenario,
the primordial background gravitational waves 
with a very blue spectrum may be produced.
This brings up the possibility that we may be able to see 
the collision epoch by a future gravitational wave
detector such as LISA.


\begin{thebibliography}{99}

\bibitem{prebigbang}
M.~Gasperini and G.~Veneziano,
Astropart.\ Phys.\  {\bf 1}, 317 (1993)
[arXiv:hep-th/9211021].
%
\bibitem{braneworld}
I.~Antoniadis, N.~Arkani-Hamed, S.~Dimopoulos and G.~R.~Dvali,
Phys.\ Lett.\ B {\bf 436}, 257 (1998)
[arXiv:hep-ph/9804398];\\
%
P.~Horava and E.~Witten,
Nucl.\ Phys.\ B {\bf 460}, 506 (1996)
[arXiv:hep-th/9510209];\\
%
P.~Horava and E.~Witten,
Nucl.\ Phys.\ B {\bf 475}, 94 (1996)
[arXiv:hep-th/9603142].
\bibitem{RS1}
L.~Randall and R.~Sundrum,
Phys.\ Rev.\ Lett.\  {\bf 83}, 3370 (1999)
[arXiv:hep-ph/9905221].
%
\bibitem{Kanno:2002py}
S.~Kanno, M.~Sasaki and J.~Soda,
 ``Born-again braneworld,''
 to appear in Prog. Teor. Phys., arXiv:hep-th/0210250.
\bibitem{kanno1}
S.~Kanno and J.~Soda,
Phys.\ Rev.\ D {\bf 66}, 083506 (2002)
[arXiv:hep-th/0207029].
\bibitem{wiseman}
T.~Wiseman,
Class.\ Quant.\ Grav.\  {\bf 19}, 3083 (2002)
[arXiv:hep-th/0201127].
\bibitem{kanno2}
S.~Kanno and J.~Soda,
Phys.\ Rev.\ D {\bf 66}, 043526 (2002)
[arXiv:hep-th/0205188].
\bibitem{kanno3}
S.~Kanno and J.~Soda,
``Braneworld effective action at low energies,''
arXiv:gr-qc/0209087;\\
%
J.~Soda and S.~Kanno,
``Holographic view of cosmological perturbations,''
arXiv:gr-qc/0209086;\\
%
T.~Shiromizu and K.~Koyama,
``Low energy effective theory for two brane systems: Covariant curvature  formulation,''
arXiv:hep-th/0210066.
\bibitem{detector} 
P. Bender et. al, LISA. Pre-Phase A Report, second edition, July 1998.
A choice of reference texts on the LISA project can be
 found at http://www.lisa.uni-hannover.de/lisapub.html.
%
\end{thebibliography}
\end{document}